\documentclass[11pt]{article}
\usepackage{amsmath,dsfont,slashed,amssymb,epsf,epsfig,amsthm,bm,graphicx}
\usepackage{color}
\definecolor{darkgreen}{rgb}{0,0.35,0}

\begin{document}

\title{ \flushright{\small CECS-PH-13/xx} \\
\center{\textbf{Local supersymmetry without SUSY partners}\footnote{This is an extended version of the article \textit{Unconventional supersymmetry and its breaking}, published in Phys. Lett. \textbf{B 735} (2014) 314.}}}
\author{\textbf{Pedro D. Alvarez $^{1,2,3}$\footnote{alvarez-at-physics-ox-ac-uk}}
\textbf{Pablo Pais$^{1,2}$\footnote{pais-at-cecs-dot-cl}}
 \and
\textbf{Jorge Zanelli$^{1,2}$\footnote{z-at-cecs-dot-cl}}\\
\textbf{\textit{$^{1}$}}{\small Centro de Estudios Cient\'{\i}ficos (CECS) Av.Arturo Prat 514, Valdivia, Chile}\\
\textbf{\textit{$^{2}$}}{\small Universidad Andr\'es Bello, Av. Rep\'ublica 440, Santiago, Chile}\\
\textbf{\textit{$^{3}$}}{\small Rudolf Peierls Centre for Theoretical Physics, University of Oxford}}
\date{}
\maketitle

\begin{abstract}
A gauge theory for a superalgebra that includes an internal gauge ($\mathcal{G}$) and local Lorentz ($\mathfrak{so}(1,D-1)$) algebras, and that for $D=4$ could describe the low energy particle phenomenology is constructed. These two symmetries are connected by fermionic supercharges. The system includes an internal gauge connection one-form $\mathbf{A}$, a spin-1/2 Dirac spinor $\psi$, the Lorentz connection $\omega^{ab}$, and the vielbein $e^a_\mu$. The connection one-form $\mathbf{A}$ is in the adjoint representation of $\mathbf{G}$, while $\psi$ is in the fundamental. In contrast to standard supersymmetry and supergravity, the metric is not a fundamental field and is in the center of the superalgebra: it is not only invariant under the internal gauge group, $\mathbf{G}$, and under Lorentz transformations, $\mathbf{SO(1,D-1)}$, but is also invariant under supersymmetry.

The distinctive features of this theory that mark the difference with standard supersymmetries are: \textbf{i)} the number of fermionic and bosonic states is not necessarily the same; \textbf{ii)} there are no superpartners with equal mass, ``bosoninos", sleptons and squarks are absent; \textbf{iii)} although  this supersymmetry originates in a local gauge theory and gravity is included, there is no gravitino; \textbf{iv)} fermions acquire mass from their coupling to the background or from higher order self-couplings, while bosons remain massless. 

In odd dimensions, the Chern-Simons form provides an action that is quasi-invariant under the entire superalgebra. In even dimensions, the Yang-Mills form $L=\kappa \langle\mathds{F}^\circledast \mathds{F}\rangle$ is the only natural option and the symmetry breaks down to $\mathbf{G}\otimes\mathbf{SO(1,D-1)}$. In four dimensions, the construction follows the Townsend - Mac Dowell-Mansouri  approach, starting with an $\mathbf{\mathfrak{osp}(4|2)} \sim \mathbf{\mathfrak{usp}(2,2|1)}$ connection. Due to the absence of $\mathbf{\mathfrak{osp}(4|2)}$-invariant traces in four dimensions, the resulting Lagrangian is only invariant under $\mathbf{\mathfrak{u}(1)} \oplus \mathbf{\mathfrak{so}(3,1)}$, and includes a Nambu--Jona-Lasinio (\textbf{NJL}) term. In this case, the Lagrangian depends on a single dimensionful parameter that fixes Newton's constant, the cosmological constant and the NJL coupling.
\end{abstract}

\section{Introduction}      
Supersymmetry (SUSY), a symmetry that unifies spacetime transformations and internal gauge symmetries, combining bosons and fermions, presents a curious paradox: On the one hand, there is a wide consensus among theoretical --and even experimental-- physicists that this unification must exist and be reflected in the particle spectrum of the standard model \cite{Martin,Murayama}. On the other, in spite of four decades of extensive search at accelerators, no evidence of SUSY has been found \cite{Chatrchyan:2011zy}, at least in its simplest form, making even some of the original proponents recant their support for the idea \cite{Shifman}. 

A distinct signal of standard SUSY would be the existence of partners that duplicate the spectrum of observed particles \cite{Wess-Bagger}. In the minimal supersymmetric scenario ($\mathcal{N}=1$ SUSY), for every lepton, quark and gauge quantum, a corresponding particle/field with identical quantum numbers but differing by $\hbar/2$ in intrinsic angular momentum would exist.\footnote{In more elaborate $\mathcal{N}\geq 2$ models, additional partners accompany every observed state.} In an unbroken supersymmetric phase these SUSY partners would have degenerate masses and since no partners have been observed even approximately reflecting this degeneracy, SUSY is believed to be severely broken at current experimental energies. 

This idea could be contrasted with Heisenberg's proposal of isospin, based on the observed slight difference in mass between the proton and the neutron, $\Delta m/ m\sim 10^{-3}$. In SUSY there is no approximate degeneracy to be explained by the symmetry; instead, there seems to be a need to explain the complete absence of a symmetry for which there are compelling mathematical arguments. The idea of unification has proven a strong guide for progress in physics ever since Newton and, therefore, the possibility of a nontrivial combination of Poincar\'e invariance and internal gauge symmetries cannot be underestimated. The fact that this is the only known mechanism for such combination would be sufficient to justify the interest in SUSY. But there is more.

On the practical side, it was soon observed that supersymmetric models exhibit improved renormalizable features, offering a mechanism to protect some parameters in the action from running under renormalization. This provides a possible natural solution for the so-called hierarchy problem \cite{Martin}, and therefore it is an interesting ingredient in all unified models of electroweak and strong interactions. Improved renormalizability also allowed for attractive ways to unify the standard model with gravity in supergravity \cite{Freund}, or perhaps in an ultimate unification scheme like superstring theory \cite{Strings}.

In spite of SUSY's undeniable appeal, the skeptic still has the right to question its logical necessity: does it solve a problem no other scheme can? If no trace of SUSY is ever experimentally found it can always be argued that SUSY breaking takes place at an energy beyond reach. But, can this be a falsifiable statement \cite{Popper}?

The origin of the mass degeneracy can be traced back to the assumption in standard (global) SUSY that all fundamental fields are in a vector representation of the supercharge $\mathds{Q}$, and that this generator commutes with the Hamiltonian. In addition, SUSY is usually expected to be defined in a globally flat, Poincar\'{e}-invariant,  spacetime, which seems unrealistic in view of the fact that we live in an evolving spacetime that need not possess a particular symmetry at any given time. If spacetime is not flat the supercharges need not commute with the Hamiltonian, lifting the mass degeneracy.\footnote{For, example, if the spacetime had constant curvature with cosmological constant $\Lambda$, the difference in mass between supersymmetrically related states  would be $\Delta m / m \sim G\sqrt{|\Lambda|}$ which, for the best current estimate is extremely small, $\sim M_{Pl}^{-2} \sqrt{10^{-120}M_{Pl}^4}\sim 10^{-60}$ \cite{deWit}.} In this sense, supersymmetry could be broken by contingent effects --spacetime not being maximally symmetric--, while the unbroken situation could be regarded as an unlikely accident, an ideal situation or an approximation to reality valid only in a small neighborhood at best.

Here we consider a theory that keeps the essence of the supersymmetric paradigm --that fermions and bosons can be combined into a nontrivial representation of a supergroup--, but which differs in three important aspects from the standard SUSY construction. First, SUSY here is an extension of the symmetries of the \textit{tangent space}. In fact, spinors, including the supercharges, are in a spin-$1/2$ representation of the Lorentz group, which is an exact invariance of the tangent space. According to the equivalence principle, any theory consistent with general relativity is invariant under Lorentz transformations acting locally on the tangent bundle, independently of the spacetime curvature \cite{Jackson-Okun}. This symmetry can be made manifest by projecting all spacetime tensors on the tangent space, allowing supersymmetry to be realized on the tangent of any curved manifold as if in Minkowski spacetime. This scheme alone, however, would still give rise to a particle spectrum with boson-fermion mass degeneracy requiring a credible supersymmetry breaking scheme.

The second point of departure from the standard global (rigid) SUSY is that we assume the fundamental fields in an adjoint representation, as parts of a connection one-form for a superalgebra, and not in a vector representation of the supergroup. In this approach, bosons and fermions are parts of the same connection, an idea that has been exploited since the mid-70s to construct supergravities \cite{Chamseddine}, and was extended to construct Chern-Simons (\textbf{CS}) gauge theories with local off-shell SUSY including gravity for all odd dimensions \cite{BTrZ,TrZ1,TrZ2,HTrZ}. This approach is particularly suited for a theory with spin-3/2 fermions, which are naturally one-forms ($\psi^\alpha=\psi^\alpha_\mu dx^\mu$), but does not seem to work for ordinary SUSY without gravitini.

The third element in the construction is the use of the vielbein to project the Clifford algebra of spinors from the tangent space onto the spacetime manifold, turning the Dirac matrices into matrix-valued one-forms. The consequence of this is that spin-3/2 fermions (gravitini) are unnecessary. Conversely, this allows reading the supersymmetry transformations as defined on the tangent space, which for all practical purposes can be taken as Minkowski spacetime.

This approach gives rise to a scenario where, as in the Standard Model, bosons are interaction carriers described by massless connection fields in the adjoint representation of the gauge algebra, while fermions are vectors under the gauge group (sections in the gauge bundle) and their currents are sources for the bosonic fields. There are no Bose-Fermi pairs, particles of different spins need not have equal masses and all fields are coupled in the standard gauge-invariant way. The theory can be defined in an arbitrarily curved background and SUSY requires the inclusion of gravity, so this model could be seen as a hybrid between standard SUSY and SUGRA. In contrast to supergravity, however, all fermions are spin-1/2 particles and no gravitini are included. In \cite{AVZ}, a model in three dimensions is constructed along the lines described above which proves the point that such an alternative to conventional SUSY exists, giving rise to a completely different scenario.

\section{Spin-1/2 fields as part of the connection}      
The main lesson from \cite{AVZ} is that the representation is crucial not only for the field content of the model, but for the dynamical relations among the different constituents. Although this is well known, it seems to have been systematically ignored by the model building industry, putting all fields by default in the vector (fundamental) representation under supersymmetric rotations. The only general exception is in supergravity, where the gravitino is expected to behave as a connection required by local SUSY transformations.

Following \cite{AVZ}, consider arranging bosonic and fermionic fields into a connection one-form as follows,
\begin{equation}
\mathds{A} \sim A^r \mathds{B}_r + \overline{\mathds{Q}} \mathbf{\Gamma} \psi + \overline{\mathds{\psi}} \mathbf{\Gamma} \mathds{Q}\; .
\end{equation} \label{connection-1}
where $A^r=A^r_\mu dx^\mu$ is a connection one-form, $\psi$ is a Dirac or Majorana spinor and the generators $\mathds{B}_a$ and $\mathds{Q}$ span a superalgebra of the form
\begin{equation}
[\mathds{B}, \mathds{B} ] \sim \mathds{B}, \quad  \{\overline{\mathds{Q}}, \mathds{Q} \} \sim \mathds{B}, \quad [\mathds{B}, \mathds{Q}] \sim \mathds{Q}, \quad [\mathds{B}, \overline{\mathds{Q}}] \sim -\overline{\mathds{Q}}.
\end{equation}

In the fermionic terms of (\ref{connection-1}) the spinor representation of the vielbein  is used
\begin{equation}
\mathbf{\Gamma} = dx^\mu \Gamma_\mu = dx^\mu e^a_\mu \Gamma_a \,
\end{equation}
where $e^a_\mu$ are the components of the vielbein and $\Gamma_a$ are the Dirac matrices defined on the tangent space, thus projecting the Clifford algebra from the tangent onto the base manifold. As is well known $e^a_\mu$ also allows to represent tensors and differential forms from the base manifold by tensors and differential forms on the tangent space. Thus, the vielbein does not play an active dynamical role, which is consistent if $e^a_\mu$ is further assumed not to transform under supersymmetry. The resulting theory will be a gauge theory of the Lorentz group by construction, with Lorentz and supersymmetry transformations locally realized on the tangent bundle.

In this framework the presence of spinors has two effects: it forces the inclusion of Lorentz symmetry in the SUSY algebra, and introduces a metric structure through the vielbein $e^a$. These two ingredients make the incorporation of gravity practically unavoidable.

\subsection{Supersymmetry}   
Under gauge transformations $\mathbb{A}$ behaves as a connection one-form, $ \mathds{A} \rightarrow \mathds{A}'= g^{-1}(\mathds{A}   +  d)g$, where $g(x)=\exp \Lambda(x)$ is an element of the gauge group and $\Lambda $ is in the algebra $\mathcal{G}$. For an infinitesimal transformation, $\delta\mathds{A} = d\Lambda +[\mathds{A},\Lambda]$. In particular, for a local supersymmetry transformation generated by $\Lambda(x)=\overline{\mathds{Q}} \epsilon -\bar{\epsilon} \mathds{Q}$, the connection changes by
\begin{equation}\label{delbigA}
\delta\mathds{A} = (\delta_\epsilon A^r)\mathds{B}_r + \overline{\mathds{Q}} \, \delta_\epsilon(\Gamma\psi) - \delta_\epsilon(\overline{\psi}\Gamma) \, \mathds{Q} \; .
\end{equation}
This translates to the component fields as
\begin{eqnarray}
\delta_\epsilon A^r&=&-i\left[\bar{\epsilon}\{\Gamma\}^r\psi+\bar{\psi}\{\Gamma\}^r \epsilon\right]  \label{delA} \\
\delta_\epsilon(e^a_\mu \Gamma_a\psi) &=&  \overrightarrow{\nabla}_\mu \epsilon , \quad
\delta_\epsilon(\overline{\psi} \Gamma_a e^a_\mu) = -\bar{\epsilon}\overleftarrow{\nabla}_\mu \; ,  \label{delpsi}
\end{eqnarray}
where $\{\Gamma\}^r$ denotes a properly (anti-)symmetrized product of Dirac matrices, and $\nabla$ is the covariant derivative  for the connection in the spin-1/2 representation of the bosonic subalgebra. Finally, it can also be checked that successive gauge transformations of $\mathbb{A}$ form a closed off-shell algebra, $[\delta_{\Lambda}, \delta_{\Delta}] \mathbb{A}= \delta_{[\Lambda, \Delta]} \mathbb{A}$.
There is no need of extra fields to close the algebra, a general feature of supersymmetric theories based on super-connections \cite{TrZ2}.

\subsection{Absence of gravitini}    
In ordinary gauge theories the metric is assumed to be invariant under the internal gauge transformations. Analogously, in this case we assume the vielbein to be invariant under supersymmetry,  $\delta_{\epsilon} e^a_\mu=0$. This allows writing (\ref{delpsi}) as
\begin{equation}
 \Gamma_\mu\delta_\epsilon \psi =  \overrightarrow{\nabla}_\mu \epsilon\; ,  \quad  \delta_\epsilon(\overline{\psi}) \Gamma_\mu = -\bar{\epsilon}\overleftarrow{\nabla}_\mu \; ,
\label{delpsi'}
\end{equation}
and therefore,
\begin{equation} \label{deltapsi}
\delta_{\epsilon}\psi=\frac{1}{D}\slashed{\nabla} \epsilon \qquad \qquad \delta_{\epsilon} \overline{\psi}=-\frac{1}{D}\bar{\epsilon} \overleftarrow{\slashed{\nabla}} \;,
\end{equation}
where $D$ is the dimension of spacetime.

The condition $\delta_{SUSY} e^a_\mu=0$ means that the metric $g_{\mu \nu}$, which is invariant under the internal gauge group and under Lorentz transformations,\footnote{The vielbein $e^a$, however, transforms as usual like a Lorentz vector.} is also invariant under supersymmetry. This means, in particular, that there is no need here to introduce gravitini, in spite of this being a supersymmetric theory in a gravitational background.

Plugging $\delta \psi$ from (\ref{deltapsi}) back into (\ref{delpsi}) yields the constraint
\begin{equation}\label{projector01}
\left( \delta_b{}^a-\frac{1}{D}\Gamma_b \Gamma^a \right) \nabla_a \epsilon=0 \; ,
\end{equation}
where $\nabla_a = E^\mu_a \nabla_\mu$ is the covariant derivative projected onto the tangent space, and $E^\mu_a$ is the inverse vielbein, $E^\mu_a e^b_\mu = \delta^b_a $. It is easy to see that $P_b{}^a \equiv \delta_b{}^a - \frac{1}{D}\Gamma_b \Gamma^a $ projects spinorial one-forms in the tangent space $\chi^\alpha_a \in 1/2 \otimes 1$, onto the spin-3/2 subspace. Its orthogonal complement, $\delta_b{}^a - P_b{}^a = \frac{1}{D}\Gamma_b \Gamma^a$ projects onto the spin 1/2 components. Consequently, (\ref{projector01}) eliminates the spin-3/2 component from $\nabla_a \epsilon$, which is consistent with the fact that no gravitini are included in the connection.

A further consistency check is that the projection operator itself is invariant under supersymmetry, both in the tangent space ($P_a{}^b$) and in the base manifold ($P_\nu{}^\mu$).\footnote{Since $e^a_\mu$ and its inverse, $E_a^\mu$, are invariant under SUSY, the projector $P_\nu{}^\mu=e^a_\nu E^{\mu}_b P_a{}^b$ also commutes with supersymmetry.}

The projection  (\ref{projector01}) is the covariant version of the constraint found by Wess and Zumino in their seminal paper \cite{WZ}. There, the spinorial parameter $\alpha(x)$ that defines a local SUSY transformation is expected to obey the constraint
\begin{equation}\label{projector0}
\left(\frac{1}{4}\Gamma_a \Gamma^b - \delta_a^b\right) \partial_b \alpha =0.
\end{equation}
The most general solution for this equation is $\partial_a \alpha =\Gamma_a \beta$. The consistency condition (integrability) for this relation implies $\partial_a \partial_b \alpha =\partial_b \partial_a \alpha$, which means that $\beta$ must be a constant spinor, and $\alpha(x)$ should be a linear function of the coordinates,
\begin{equation*}
\alpha(x) = \alpha_0 +x^a \Gamma_a \beta \, .
\end{equation*}
This has two important consequences. First, supersymmetry in the Wess-Zumino model is not a gauge symmetry, described by arbitrary local functions, but it is a rigid transformation. The SUSY transformations are parametrized by two constant spinors $\alpha_0$ and $\beta$ and therefore the mixing between fermions and bosons everywhere in spacetime depends on the values of these two constant spinors. Second, the spacetime manifold must be flat (four-dimensional) Minkowski space, because only in a flat manifold the combination $x^a \Gamma_a$ has an unambiguous meaning.

In our construction, the general solution of (\ref{projector01}) is given by $\nabla_a \epsilon=\Gamma_a \beta$, for an arbitrary spinor $\beta$, and the consistency relation is
\begin{equation}
[\nabla_a \nabla_b -\nabla_b \nabla_a] \epsilon = \left[ \Gamma_b \Gamma_a- \Gamma_a \Gamma_b \right] \beta . \label{epsilon-beta}
\end{equation}
The left hand side is an algebraic expression involving the curvature components of the bosonic gauge connections; (\ref{epsilon-beta}) establishes an algebraic relation at each point of the background between $\epsilon$ and some arbitrary $\beta$. The number of independent globally defined solutions of equation (\ref{projector01}) depends on the gauge curvatures and on possible topological obstructions. In a typical experimental setting in accelerators, however, the curvatures are negligible in the region where the experiment are carried out. The relevant regions in those cases are huge compared with the quantum wavelength of the particles involved, but at the same time are extremely small compared with the local radius of curvature of spacetime and, to a good approximation the curvature can be safely assumed to vanish. Then $\beta$ can be taken equal to zero and  $\epsilon$ approximates a Killing spinor.

It is not so obvious how this equation is solved in backgrounds not continuously connected to the globally $\mathds{F}=0$ configuration, which might lead to topological obstructions that break supersymmetry. In this sense, this type of supersymmetry may be as fragile as a standard one that assumes Minkowski spacetime.  However, the vanishing gauge curvature ($\mathds{F}=0$) is a generic property of the odd dimensional Chern-Simons vacua, in which case the ground states can be expected to be supersymmetric. In even dimensions, however, this is no longer true: the ``vacua" need not have vanishing gauge curvature, but in that case the action itself is not locally supersymmetric due to the nonexistence of a (super) gauge-invariant action in even dimensions.

In the construction outlined here the incorporation of supersymmetry in the gauge algebra strongly restricts the field content of the theory. In particular, the resulting theory requires the inclusion of a soldering form  $e^a_\mu$ invariant under supersymmetry, a Lorentz connection $\omega^a{}_{b\mu}$, and an internal gauge connection $A^K_\mu$, apart from the spin-1/2 field, charged with respect to the internal gauge interaction. The dynamics of these fields and the way they couple with each other is dictated by the connection which depends critically on the spacetime dimension, and the Lagrangian that is used. Here we consider using the CS form for odd dimensions and the YM form in even dimensions, but of course other options exist.

\section{Constructing Lagrangian $D$-forms}      
The dynamical features of a system described in terms of these fields should be obtained from a Lagrangian $L(\mathds{A})$ that is expected to be either an invariant or quasi-invariant\footnote{A function $f(\mathds{A})$ is quasi-invariant if under a gauge transformation it changes by a locally exact form, $\delta f=d \phi$.}  polynomial in $\mathds{A}$ and $d\mathds{A}$. The associated curvature $\mathds{F}=d\mathds{A}+\mathds{A}\wedge \mathds{A}$ (field strength), is a tensor under gauge transformations in the adjoint representation, $\mathds{F} \rightarrow \mathds{F}'= g^{-1}\mathds{F} g$. The obvious invariant choice in even dimensions\footnote{Exterior (wedge) products of differential forms will be implicitly assumed throughout.}
\begin{equation}
P_{2n}= \left\langle \mathds{F} \cdots \mathds{F} \right\rangle  \; ,
\label{ChernClass}
\end{equation}
where $\left\langle \cdots \right\rangle$ is a (super) trace in the Lie algebra, is an invariant polynomial 2n-form. However, this is a topological invariant and not a suitable Lagrangian. In fact, the Chern-Weil theorem asserts that any invariant polynomial of this form is necessarily closed, $dP_{2n}=0$, and therefore it is \textit{locally} an exact form: $P_{2n}=dC_{2n-1}$ \cite{Nakahara}. This means that its variations --under appropriate boundary conditions-- identically  vanish, or are just a boundary term, while the dynamics in the bulk remains arbitrary. Thus, in particular, there are no Lagrangians $L(\mathds{F})$ constructed using only exterior products, invariant under the entire gauge group; the Euler-Lagrange equations for such ``invariant Lagrangians" would have the trivial form $0=0$. In order to have dynamics in even dimensions one must give up gauge invariance under the full gauge group.

This leaves essentially two possibilities for a viable Lagrangian: i) instead of being simply invariant, it can be \textit{quasi-invariant} --that is, $L(\mathds{A})$ changes by a total derivative under gauge transformations--; or ii) it can be invariant under a \textit{proper subgroup} of the gauge group.

The first case corresponds to Lagrangians defined by CS forms\footnote{CS theories have been extensively discussed in the physics literature, starting with the pioneering works of Cremmer, Julia and Sherk \cite{CJS}, Schoenfeld \cite{Schonfeld}, and Deser, Jackiw and Templeton \cite{DJT}. For a recent review, see \cite{Review}.} that define dynamical theories in odd dimensions. For example, given a Lie algebra-valued connection $A$ in $2n+1$ dimensions, the CS form is naturally defined,
\begin{equation}
L^{CS}_{2n+1} \equiv \mathcal{C}(\mathds{A}) = \left\langle \mathds{A} d\mathds{A}^n +... \right\rangle \; ,
\end{equation}
where the supertrace $\left\langle \cdots \right\rangle$ is invariant under the entire gauge group. Under gauge transformations continuously connected to the identity, the CS form changes by a boundary term by construction, $\delta \mathcal{C}= d\Omega$. Thus, in odd dimensions the problem reduces to find the invariant bracket $\left\langle \cdots \right\rangle$.

The second case occurs if the form (\ref{ChernClass}) is constructed with a symmetric trace  $\left\langle \cdots \right\rangle$ that is not invariant under the entire gauge symmetry group, but under a subgroup of it. This case is the only alternative in even dimensions and corresponds to the approach taken by Mac Dowell and Mansouri \cite{MD-M}, and by Townsend \cite{Townsend} to construct a four-dimensional (super)gravity out of a superalgebra for the (super-)AdS symmetry. Those authors found that although the fields could be described by an $SO(3,2)$ (AdS$_4$) connection, the four-dimensional action could be at most invariant under the Lorentz group ($SO(3,1)$-invariant).

In all dimensions, YM Lagrangians can be constructed, provided the spacetime is equipped with a metric structure with which the Hodge dual of $\mathds{F}$ is defined. Thus, we tentatively define
\begin{equation} \label{YM}
L^{YM} = -\frac{1}{4} Str\left[\mathds{F}{}\wedge ^{\circledast}\mathds{F} \right] \;,
\end{equation}
where $^\circledast \mathds{F}$ is the dual of $\mathds{F}$. The metric structure required by this construction is provided by the soldering form $e^a$.

\subsection{Three-dimensional example}   
In three dimensions, the construction outlined above leads to the model discussed in \cite{AVZ}. We summarize the results here to illustrate the idea, further details can be found in that reference. The connection (\ref{connection-1}) takes the form
\begin{equation}
\mathbb{A}=A \mathbb{K} + \overline{\mathbb{Q}}_{\beta} (\mathbf{\Gamma})^{\beta}{}_{\alpha} \psi^{\alpha} + \overline{\psi}_{\alpha}  (\mathbf{\Gamma})^{\beta}{}_{\alpha} \mathbb{Q}^{\beta} +\omega^a \mathbb{ J}_{a}  ,
\label{A3}
\end{equation}
where $\mathbb{K}$, $\mathbb{Q}$, $\overline{\mathbb{Q}}$, and $\mathbb{J}$ are the $U(1)$ generators,\footnote{These results can be extended with very small modifications to include $SU(2)$ instead of $U(1)$ \cite{APRSZ}.} supersymmetry and Lorentz transformations in 2+1 dimensions, respectively.  Here $\omega^a_{\mu}=\frac{1}{2}\epsilon^a{}_{bc}\omega^{bc}_{\mu}$ is the Lorentz connection.
The CS 3-form provides a Lagrangian for the connection $\mathbb{A}$ without additional ingredients,
\begin{equation} \nonumber
L=\langle \mathbb{A}d\mathbb{A} + \frac{2}{3} \mathbb{A}^3 \rangle \, .
\end{equation}
In the standard representation for $\Gamma$ matrices and supertrace, the Lagrangian reads
\begin{eqnarray}
L&=& 2 A d A + \frac{1}{4}[\omega^a{}_b d \omega^b{}_a +\frac{2}{3} \omega^a{}_b \omega^b{}_c \omega^c{}_a ] - 2\overline{\psi} \psi e^aT_a \nonumber \\
&& + 2 \overline{\psi} (/\hspace{-8pt}\overleftarrow{\partial}-/\hspace{-8pt} \overrightarrow{\partial} +  2i /\hspace{-7pt}  A + \frac{1}{2} \gamma^a /\hspace{-7pt} \omega_{ab} \gamma^b) \psi |e| d^3x , \label{L}
\end{eqnarray}
where $|e|=$det$[e^a{}_{\mu}]=\sqrt{-g}$, and $T^a=de^a+\omega^a{}_{b} e^b$ is the torsion 2-form. This is a standard Lagrangian for a Dirac field minimally coupled to CS electrodynamics in a gravitational background \cite{Weyl}. The system is invariant under local $U(1)$ and $SO(2,1)$ transformations. It may be surprising that this rather ordinary-looking system is obtained as a gauge theory for the $osp(2|2)$ superalgebra. Although this supersymmetry is local and contains 2+1 gravity, there is no gauging of local translations and hence, no gravitino is required.

The field equations for this system are
\begin{eqnarray}
\label{F=psi2}
&& F_{\mu \nu} =\epsilon_{\mu \nu \lambda} j^{\lambda} \\
\label{R=psi2}
 & &R^{ab} = 2\overline{\psi} \psi e^a e^b \\
\label{Dirac}
& & [\slashed{\partial}   - i  \slashed{A} + \mu - \frac{1}{4}\Gamma^a \slashed{\omega}_{ab} \Gamma^b +     \frac{1}{2|e|}\partial_\mu(|e|E^\mu_a \Gamma^a) ] \psi =0 \, ,
\end{eqnarray}
where $j^{\lambda}=-i\overline{\psi} \Gamma^{\lambda} \psi |e|$, is the electric current density of a charged spin 1/2 field, and $|e|\mu d^3x \equiv e^aT_a$. Since $R^{ab}e_b =DT^a$ and $e^a e^b e_b \equiv 0$, (\ref{R=psi2}) implies that the torsion is covariantly constant, $DT^a=0$ and therefore $\mu$ must be an (arbitrary) constant that can be identified with the fermion mass.

The matter-free  configurations $\psi=0$ imply $F=0=R^{ab}$ and, as shown in \cite{AVZ} this corresponds to manifold whose local geometry has constant torsion and constant negative Riemannian curvature. These anti-de Sitter spaces include rotating and magnetically charged BTZ black holes and some naked conical singularities corresponding to rotating and charged point sources \cite{Miskovic-Z,EGMZ}. For some values of mass ($M$), angular momentum ($J$) and magnetic charge ($q$), these configurations are BPS states and therefore correspond to stable supersymmetric vacua. Moreover, for arbitrary values of $M$, $J$ and $q$ these configurations are locally AdS-flat and therefore satisfy the consistency conditions (\ref{epsilon-beta}) for $\beta=0$.

In addition to these formal properties, the Lagrangian (\ref{L}) describes the propagation dynamics of carriers of electric charge in graphene in the long wavelength limit near the Dirac point \cite{VKG,MSZ}. In fact, one of the salient features of the graphene system seems to be its conformal symmetry $\psi \rightarrow \Omega \psi$, $e^a \rightarrow \Omega^{-1} e^a$ \cite{Cvetic-Gibbons,Iorio}, which in our model is a natural consequence of the construction.

\subsection{Four-dimensional action}    
Let us now see how would this construction operate in four dimensions. In Appendix B, the simplest SUSY in four dimensional space containing $U(1) \times SO(3,1)$ is presented. This is the $\mathbf{\mathfrak{osp}(4|2)} \sim \mathbf{\mathfrak{usp}(2,2|1)}$ superalgebra and includes the (A)dS$_4$ generators $\mathds{J}_a$ and $\mathds{J}_{ab}$, the complex supercharge $\mathds{Q}^\alpha$ in a spin 1/2 representation, and the $U(1)$ generator $\mathds{K}$. 
\subsubsection{SUSY algebra, connection and curvature}  
The essential anticommutator of the superalgebra is
\begin{equation}
\{\mathds{Q}^\alpha , \overline{\mathds{Q}}_\beta \}=-i(\Gamma^a)^{\alpha}_{\beta} \mathds{J}_a +\frac{i}{2} (\Gamma^{ab})^{\alpha}_{\beta}\mathds{J}_{ab} -\delta^{\alpha}_{\beta}\mathds{K} \, , \label{Q-barQ}
\end{equation}
together with the trivial anticommutators $\{\overline{\mathds{Q}}_\alpha,\overline{\mathds{Q}}_\beta\}=0 = \{\mathds{Q}^\alpha,\mathds{Q}^\beta\}$. An explicit $6\times 6$ representation for the supercharges is
\begin{equation}
(\mathds{Q}^\alpha)^A_{\,B} = -\frac{i}{s}(\delta^A_5 \delta^\alpha_B+ C^{\alpha A} \delta^6_B) , \quad
(\overline{\mathds{Q}}_\alpha)^A_{\,B} = \delta^A_\alpha \delta^5_B+\delta^A_6 C_{\alpha B}\, ,
\end{equation}
where $s^2 = -1$ corresponds to de Sitter, and $s^2 = 1$ to anti-de Sitter. Here $C_{\alpha \beta}=-C_{\beta \alpha}$ is the conjugation matrix, $C^{\alpha \beta}$ is its inverse\footnote{The indices $A,B=1,...,6$ combine both spinor indices ($\alpha, \beta = 1,...,4$) and those of a two-dimensional representation ($r=5,6$) of $U(1)$, i.e., $A=(\alpha, r)$.}.
In this representation, the $U(1)$ and AdS generators are
\begin{equation}
 {(\mathds{K})^A}_B = i(\delta^A_5 \delta^5_B-\delta^A_6 \delta^6_B), \quad
 {(\mathds{J}_a)^A}_B =\frac{1}{2}{(\Gamma_a)^\alpha}_\beta \delta^A_\alpha \delta ^\beta_B , \quad
\quad {(\mathds{J}_{ab})^A}_B =\frac{1}{2}{(\Gamma_{ab})^\alpha}_\beta \delta^A_\alpha \delta ^\beta_B\,
\end{equation}

The connection can be written as
\begin{equation}
\mathds{A} = A \mathds{K} + \overline{\mathds{Q}} \mathbf{\Gamma} \psi + \overline{\psi}\mathbf{\Gamma} \mathds{Q} + f^a\mathds{J}_a + \frac{1}{2}\omega^{ab} \mathds{J}_{ab} , \label{connection}
\end{equation}
 where $A=A_\mu dx^\mu$, $\mathbf{\Gamma}=\Gamma_a e^a_\mu dx^\mu$, $f^a=f^a_\mu dx^\mu$ and $\omega^{ab} =\omega^{ab}_\mu dx^\mu$ are 1-form fields (spinorial indices omitted). The curvature $\mathds{F}= d\mathds{A} + \mathds{A} \mathds{A}$ takes the form $\mathds{F} =F_0 \mathds{K} + \overline{\mathds{Q}}_{\alpha}\mathcal{F}^{\alpha} + \overline{\mathcal{F}}_{\alpha}\mathds{Q}^{\alpha} + F^a\mathds{J}_a + \frac{1}{2}F^{ab} \mathds{J}_{ab}$,  where
\begin{eqnarray}
F_0 &=& F - \overline{\psi} \slashed{e}\slashed{e} \psi \\ \label{Ffirst}
\mathcal{F} &=& \nabla (\slashed{e} \psi)\\ \label{calF}
\overline{\mathcal{F}}&=& (\overline{\psi} \slashed{e} ) \overleftarrow{\nabla} \\ \label{calFbar}
F^a &=& Df^a - \frac{i}{s}\overline{\psi}\slashed{e}\Gamma^a \slashed{e}\psi \\
F^{ab} &=&R^{ab} + s^2 f^af^b + i\overline{\psi}\slashed{e} \Gamma^{ab} \slashed{e}\psi \,,\label{Flast}
\end{eqnarray}
Here $F=dA$, $Df^a=df^a+ \omega^a{}_b f^b$, and $R^a{}_b =d\omega^a{}_b + \omega^a{}_c \omega^c{}_b$. We have also used the notation $\slashed{f}= \Gamma_a f^a$, $\slashed{e}= \Gamma_a e^a \equiv \mathbf{\Gamma}$, and $\slashed{\omega}= \frac{1}{2}\Gamma_{ab}\omega^{ab}$. The operators $\nabla \equiv  \left[d - iA + \frac{s}{2}\slashed{f}  + \frac{1}{2}\slashed{\omega} \right]$  is the covariant derivative in the spin-1/2 representation (and $ [-\overleftarrow{d} -iA + \frac{s}{2}\slashed{f} + \frac{1}{2} \slashed{\omega} ] \equiv \overleftarrow{\nabla}$).

\subsubsection{Supersymmetry transformations}   
Under a supersymmetry transformation generated by $\Lambda=\overline{\mathds{Q}}\epsilon - \overline{\epsilon}\mathds{Q}$, the connection $\mathds{A}$ changes by $\delta\mathds{A}=d\Lambda+[\mathds{A},\Lambda]$.  Using the (anti-)commutation relations of the superalgebra, one finds
\begin{eqnarray}\label{delta-susy-A}
\delta A_\mu&=&-\left(\overline{\epsilon}\Gamma_\mu\psi +\overline{\psi}\Gamma_\mu\epsilon \right)\\
\label{delta-susy-f}
\delta f^a &=& -\frac{i}{s} \left(\overline{\epsilon}\Gamma^a\slashed{e}\psi + \overline{\psi}\slashed{e} \Gamma^a\epsilon \right)\\
\label{delta-susy-omega}
\delta\omega^{ab}&=& i\left(\overline{\epsilon}\Gamma^{ab}\slashed{e}\psi + \overline{\psi}\slashed{e} \Gamma^{ab}\epsilon \right)\\
\label{delta-susy-psi}
 \delta\left[\Gamma_\mu\psi \right]&=&\left[\partial_\mu -i A_\mu + \frac{s}{2} f^a_\mu \Gamma_a + \frac{1}{4}\omega^{ab}_\mu \Gamma_{ab} \right]\epsilon \equiv \nabla_\mu \epsilon \,.
\end{eqnarray}
As discussed above, using $\delta e^a=0=\delta \Gamma_\mu$ in (\ref{delta-susy-psi}) implies  $\delta\psi = \frac{1}{4}\Gamma^\mu\nabla_\mu \epsilon$, and the consistency condition $\left[\delta^\mu_\nu - \frac{1}{4}\Gamma_\nu \Gamma^\mu \right]\nabla_\mu \epsilon=0$ eliminates the spin-3/2 part.

\subsubsection{Invariant Hodge supertrace}     
Starting from the connection (\ref{connection}), one can construct an action of the YM type. The Lagrangian is a four-form quadratic in curvature,
\begin{equation}
L=\kappa \langle\mathds{F}  ^\circledast \mathds{F}  \rangle,
\label{Fsquare}
\end{equation}
where $^\circledast \mathds{F} $ stands for the dual of $\mathds{F}$ with $(^\circledast)^2=-1$ in the Lorentzian signature. Here we take duality as the Hodge dual $(*)$ in the spacetime, the $\Gamma_5$-conjugate in spinor indices, and the dual in the AdS algebra, to wit,
\begin{equation}
^\circledast \mathds{F}   = *F_0 \mathds{K} + (\overline{\mathds{Q}})_\alpha (\Gamma_5 \mathcal{F})^\alpha + (\overline{\mathcal{F}})_\alpha ( \Gamma_5 \mathds{Q})^\alpha + \mathbf{\Upsilon}\left[ F^a\mathds{J}_a + \frac{1}{2}F^{ab} \mathds{J}_{ab}\right]\,.
\end{equation}
In the  $6\times 6$ representation, $(\mathbf{\Upsilon})^A{}_B=(\Gamma_5)^{\alpha}{}_{\beta} \delta^{\beta}_B \delta^A_{\alpha}$, or
\begin{equation}
\mathbf{\Upsilon}=\left[\begin{array}{c|cc}
\Gamma_5 & \begin{array}{c}
 0\\
 0\\
 0\\
 0
 \end{array} & \begin{array}{c}
 0\\
 0\\
 0\\
 0
 \end{array}\\
\hline
0\; 0 \; 0\; 0\; & 0 & 0 \\
0\; 0 \; 0\; 0\; & 0 & 0
\end{array}\right] \,
\end{equation}
The three dualities square to minus the identity in their respective subspaces, $(\ast)^2 =  (\Gamma_5)^2=(\mathbf{\Upsilon})^2=  -1$ \footnote{This choice of the dual operator $\circledast$ ensures that it produces the right kinetic terms for the Maxwell filed, the gravitational action and the spinor.}. Since $\mathbf{\Upsilon}$ commutes with $\mathds{K}$ and $\mathds{J}_{ab}$, but not with $\mathds{J}_a$ or $\mathds{Q}^\alpha_i$, the resulting quadratic form (\ref{Fsquare}) is invariant under $SO(3,1)\times U(1)$, the only remaining symmetry of the action out of the full AdS supersymmetry (\ref{Q-barQ}).

\subsubsection{4D Lagrangian}        
The nonvanishing supertraces, bilinear in the generators that appear in $L$, are
\begin{equation}
\langle \mathds{K} \mathds{K} \rangle = 2,  \qquad \langle \mathds{Q}^\alpha \overline{\mathds{Q}}_\beta \rangle  = 2i \delta^{\beta}_{\alpha}= - \langle  \overline{\mathds{Q}}_\alpha \mathds{Q}^\beta \rangle,  \qquad \langle \mathds{J}_{ab}\mathbf{\Upsilon}\mathds{J}_{cd} \rangle = \epsilon_{abcd}\, ,
\end{equation}
and therefore,
\begin{equation}\label{action0}
\langle\mathds{F}  ^\circledast \mathds{F}\rangle = 2 F_0 \ast F_0 +4i \overline{\mathcal{F}}_\alpha (\Gamma_5)^{\alpha}_{\beta} \mathcal{F}^\beta + \frac{1}{4}\epsilon_{abcd} F^{ab} F^{cd} \, .
\end{equation}

From (\ref{Ffirst}) and (\ref{calFbar}) it is clear that the covariant derivative acts on the components $\xi^\alpha_\mu \equiv \Gamma_\mu\psi^\alpha$ which are in the kernel of the spin-3/2 projector, $P_\mu{}^\nu\Gamma_\nu\psi=0$. The second term of the r.h.s. of (\ref{action0}) contains only covariant derivatives in the spin-1/2 representation, so we can safely assume that no dynamical channels are available to switch on a spin-3/2 excitation.

Using the conventions in Appendix A, the Lagrangian can be expressed as
\begin{equation} \label{Lagrangian00}
\begin{array}{ll}
L=-\frac{1}{4}\langle\mathds{F} ^\circledast \mathds{F}\rangle =& -\frac{1}{4}F_{\mu\nu}F^{\mu\nu}  |e|d^4x -\frac{1}{16}\epsilon_{abcd} (R^{ab} + s^2 f^a f^b) (R^{cd} + s^2 f^c f^d)  \\
 & \\
&  +\frac{i}{2} s \overline{\psi} \left[ \overleftarrow{D} \Gamma_5\slashed{e}\slashed{f} \slashed{e}  + \Gamma_5\slashed{e}\slashed{f} \slashed{e}\overrightarrow{D}\right]\psi + \frac{i}{2} s \overline{\psi} \left[\Gamma_5(\slashed{T}\slashed{f} \slashed{e} - \slashed{e} \slashed{f}\slashed{T}) \right] \psi \\
& \\
& - \frac{i}{2}s^2 \overline{\psi} \Gamma_5 \slashed{e} \slashed{f}\slashed{f} \slashed{e}\psi   +12\left[(\overline{\psi} \Gamma_5 \psi)^2 - (\overline{\psi} \psi)^2 \right] |e|d^4x \; ,
\end{array}
\end{equation}
where $\overrightarrow{D}\psi \equiv (d - i A  + \frac{1}{2}\slashed{\omega})\psi$, and $\bar{\psi} \overleftarrow{D} \equiv \bar{\psi}(\overleftarrow{d} + i A  - \frac{1}{2} \slashed{\omega})$. The quartic fermionic expression is the Nambu--Jona-Lasinio (\textbf{NJL}) term, $g[(\overline{\psi} \psi)^2 - (\overline{\psi} \Gamma_5 \psi)^2 ]$.

The field $f^a$ is undifferentiated and therefore its field equation can --in principle-- be algebraically solved and substituted back in the action. Since $f$ is a connection component, this means that the invariance of the theory under local AdS boosts is frozen, which is consistent with the fact that the action is not really invariant under local AdS boosts. The same is true about the vierbein $e^a$ in the first order formulation of four-dimensional gravity \cite{Review}: in that case, the torsion equation can be algebraically solved for the spin connection, underscoring the fact that 4D gravity has local $SO(3,1)$ invariance, and no $SO(3,2)$, $SO(4,1)$, or $ISO(3,1)$ symmetry. 

The tensor character of $f^a$ and $e^a$ is the same, and it was suggested in \cite{Townsend} that they should be proportional, $f^a= \mu e^a$, where $\mu$ is a constant with dimension of (length)$^{-1}$. This choice eliminates parity-violating terms from the Lagrangian, so that in the absence of parity changing interactions, this sector remains self-contained, but it might be of interest to see the consequences of relaxing this condition and to explore, in particular, whether this could lead to new phenomena in conflict with observations. If one follows the proposal in \cite{Townsend} and using the identities (I, II) in the Appendix A the Lagrangian is found to be
\begin{equation}\label{Lagrangian02}
\begin{array}{ll}
L &= -\frac{1}{4} F_{\mu \nu}F^{\mu \nu} |e| d^4x -\frac{1}{16}\epsilon_{abcd} (R^{ab} + s^2 \mu^2 e^a e^b) (R^{cd} + s^2 \mu^2 e^c e^d) \\
 & \\
& \quad -\frac{i}{2}s \mu\left[(\overline{\psi} \overleftarrow{D}) \Gamma^a \psi - \overline{\psi} \Gamma^a (\overrightarrow{D} \psi)\right] \epsilon_{abcd}e^b e^c e^d + 2i s \mu \overline{\psi} \Gamma_5\Gamma_a\psi (T_b e^b)e^a \\
 & \\
& \quad - \frac{i}{2} s^2 \mu^2 \overline{\psi} \psi \epsilon_{abcd}e^a e^b e^c e^d + 12\left[(\overline{\psi} \Gamma_5\psi)^2 - (\overline{\psi}\psi)^2\right] |e| d^4x\, .
\end{array}
\end{equation}

In standard units, $\hbar=c=1$, $\mu$ has units of mass. The spin-1/2 field with the right physical dimensions is $\psi_{physical} =\sqrt{6\mu} \psi$, where we have included a factor $\sqrt{6}$ for later convenience. Rewriting the Lagrangian in this convention, one obtains
\begin{equation}\label{Lagrangian04}
L =\left[L_F  + L_{EM} \right] \sqrt{|g|} d^4x -\frac{1}{16}\epsilon_{abcd} \left[R^{ab} + s^2 \mu^2e^a e^b \right] \left[R^{cd} + s^2 \mu^2 e^c e^d\right] ,
\end{equation}
where $L_{EM}=-\frac{1}{4}F_{\mu \nu}F^{\mu \nu}$ and the fermionic Lagrangian is
\begin{equation}
L_F = - \frac{i}{2}s\left[\overline{\psi}(\overleftarrow{\slashed{\nabla}} - \overrightarrow{\slashed{\nabla}})\psi +4\mu \overline{\psi} \psi \right] - i s t^\mu \overline{\psi}\Gamma_5 \Gamma_\mu \psi - \frac{1}{3\mu^2}\left[(\overline{\psi} \psi)^2 - (\overline{\psi}\Gamma_5 \psi)^2 \right] \; .
\end{equation}
Here $\overrightarrow{\slashed{\nabla}} \psi = (\slashed{\partial} -i\slashed{A} +\frac{1}{2}\slashed{\omega} ) \psi $, and $\overline{\psi}\overleftarrow{\slashed{\nabla}} = \overline{\psi}(\overleftarrow{\slashed{\partial}} +i\slashed{A} -\frac{1}{2}\slashed{\omega} )$, are the covariant derivatives for the connection of the [(anti-)de Sitter]$\times U(1)$ gauge group in the spinorial representation, and following \cite{Carroll}, we defined $t^\mu \equiv -\frac{1}{3!}\varepsilon^{\mu \nu \rho \tau} e^{a}_{\nu}T_{a \rho \tau}|e|$.  The correct sign of Newton's constant in (\ref{Lagrangian04}) is obtained for $s^2=-1$, that is, for the de Sitter group (see Appendix B).

\subsubsection{Field equations}   
Varying the action (\ref{Lagrangian02}) with respect to the dynamical fields yields the following (we take the de Sitter signature, $s^2=-1$):

\begin{eqnarray}
\delta A_\nu: &  \partial_{\mu}F^{\mu\nu} + i\overline{\psi}\Gamma^{\nu}\psi=0 \label{F-mu-nu-2} \\
\delta \omega^{ab}{}_\mu: & \overline{\psi} \Gamma_{ab}{}^c \psi E_c^{\mu}  + 3\mu^2\left[ E_a^\nu E_b^\lambda E_c^\mu + 2E_a^\mu E_b^\nu E_c^\lambda \right] T^c_{\nu \lambda}=0 \label{T} \\
\delta\overline{\psi}_{\alpha}: & -\overrightarrow{\slashed{\nabla}}\psi + 2i\mu \psi  + \Gamma_5 \Gamma_\mu \psi t^\mu  + \frac{2}{3\mu^2}\left[(\bar{\psi}\Gamma_5 \psi)\Gamma_5 - (\bar{\psi}\psi) \right]\psi =0 \label{psibar} \\
\delta e^a: &  \epsilon_{abcd}(R^{bc} - \mu^2 e^b e^c )e^d= \tau_a,  \label{t-mu-nu}
\end{eqnarray}
where $\tau_{a}$ is the stress-energy three-form, defined by $\delta\left(|e|[L_{F}+L_{EM}]\right)=\delta e^{a}\wedge \tau_{a}$.
From the second equation it follows that $T^c_{\mu \nu}E^\nu_c=0$, which means that torsion is determined by the local presence of fermions.
\begin{equation}\label{Torsion}
T^a_{\mu \nu} = \frac{-i}{ 3s \mu^2} \overline{\psi}\Gamma^a{}_{bc}\psi e^b_\mu e^c_\nu \; .
\end{equation}
Contracting the third equation with $\overline{\psi}_{\alpha}$ and its conjugate with $\psi^{\alpha}$, gives
\begin{equation}\label{ecuacion_2}
\overline{\psi}_{\alpha}\frac{\delta L}{\delta\overline{\psi}_{\alpha}}-\frac{\delta L}{\delta\psi^{\alpha}}\psi^{\alpha} = \partial_\mu(is \sqrt{|g|}\overline{\psi}\Gamma^{\mu}\psi)d^4x=0 \; ,
\end{equation}
which expresses the conservation of electric charge and coincides with the current conservation condition obtained from (\ref{F-mu-nu-2}).

\section{Discussion}   
\textbf{A.} In four dimensions the kinetic terms have the right form --second order Maxwell and Einstein-Hilbert terms for bosons, and first order Dirac term for fermions--, and the couplings are also the right ones to guarantee the gauge invariance of the action. The symmetry algebra, however, is not that of the connection ($\mathfrak{osp}(4|2)$), but $\mathfrak{u}(1) \times \mathfrak{so}(3,1) \subseteq \mathfrak{osp}(4|2)$. As we saw, the reduction of symmetry is due to the lack of an $OSP(4|2)$-invariant trace $\left\langle ...,...\right\rangle$, to define an invariant action \cite{Wise}. As noted by Townsend \cite{Townsend} and Mac Dowell and Mansouri \cite{MD-M}, there is not even an $SO(3,2) \subset OSP(4|2)$-invariant trace that could be used to build a local AdS-invariant gravity action in four dimensions.

The root of this obstruction can be found in the Chern-Weil theorem, which states that any locally $G$-invariant four-forms constructed out of a $G$-connection must be a characteristic class \cite{Nakahara}. Therefore, a nontrivial Lagrangian must necessarily break $G$-invariance down to a smaller group $H \subseteq G$. It can be seen that $H$ is the isotropy (or stability) subgroup of the invariance group of the tangent manifold ($G$) \cite{Catren}. In the case case at hand, $H=U(1)\times SO(D-1,1)$.\footnote{One possibility is for this symmetry breaking to emerge from the dimensional reduction to $D=4$ from a fully gauge-invariant CS theory, based on a transgression form in $D'=2n+1>4$ \cite{AWZ,MPW}.}\\

\noindent
\textbf{B.}  The fermionic Lagrangian $L_F$ describes an electrically charged spin-1/2 field, minimally coupled to the electromagnetic field and to the spacetime background, plus NJL couplings in the four-dimensional case. The coupling to torsion is not a new feature of this model but, as noted long ago by H. Weyl \cite{Weyl}, it is present whenever the Dirac equation is written in a curved spacetime with torsion. The NJL term in the four-dimensional theory is the main modification predicted by this model.\footnote{If instead of the $U(1)$ gauge group, one had considered $SU(2)$ or $SU(3)$, NJL term would have been of the form $C_{abcd}[(\overline{\psi}^a \psi^b)(\overline{\psi}^c\psi^d)- (\overline{\psi}^a\Gamma_5\psi^b)(\overline{\psi}^c \Gamma_5\psi^d)]$, where $C_{abcd}$ is an invariant tensor in the algebra.}\\

\noindent
\textbf{C.} A feature of supersymmetry obtained with this construction is the fact that the action has no fundamental dimensionful constants, and that the theory is by construction invariant under local Weyl transformations $e^a \rightarrow \rho e^a$ and $\psi \rightarrow \rho^{-1} \psi$. However, if one wants to fit a vielbein $e^a$ with dimensions of length ($\ell$) and a fermion $\psi$ with dimensions $\ell^{(1-D)/2}$ in a dimensionless connection, it is necessary to bring in an arbitrary dimensionful constant ($\mu \sim \ell^{-1}$). In three dimensions, this appears in the integration constant for $DT^a=0$ ($T^a=\mu \epsilon^{abc}e_b e_c$); in four dimensions the scale comes with the identification between the vielbein and the part of the connection related to the symmetry that is explicitly broken by the Yang-Mills form (\ref{YM}), the AdS boosts ($f^a=\mu e^a$).

It is the dimensionful constant $\mu$ which fixes all remaining parameters of the theory. In three dimensions, the electric charge and Newton's constant are $e=1$ and $G=1 $, respectively; the cosmological constant is $\Lambda=-\mu^2$, and the fermion mass, $m=\mu$. In four dimensions, the electric charge is $e=1$, Newton's constant is $G=-s^2 (4\pi \mu^2)^{-1}$, the cosmological constant is $\Lambda=-s^2 \mu^2$, and the Nambu--Jona-Lasinio coupling $g=(3\mu)^{-2}$.\\

\noindent
\textbf{D.} Both the four-fermion NJL coupling and the gravitational action are perturbatively non-renormalizable. This  strongly suggests that the whole system should be considered as a low energy effective model and not as a fully consistent quantum theory. However, the parameters of the theory are so tightly constrained that it is conceivable that the two evils may cancel each other. The exploration of this problem, however, lies well beyond the scope of this work.

The NJL term provides a mechanism for spontaneous symmetry breaking that gives mass to the fermionic excitations in superconductivity, originally proposed as a way to describe massive excitations in strong interactions \cite{Nambu}, and is important in the study finite temperature and density effects in QCD \cite{Loewe:2013zaa}. The value of the fermion mass $m$ is produced through the gap equation for a cut-off $\mathcal{M}$,
\begin{equation}
\frac{m^2}{\mathcal{M}^2}\log\left[1+\frac{\mathcal{M}^2}{m^2} \right]= 1 - \frac{2\pi^2}{g\mathcal{M}^2}.
\end{equation}
For $m=m_e \approx 0.5$ MeV  and $\mathcal{M}= M_{\text{Planck}}=G^{-1/2}\approx 2.5\times 10^{22} m_e$, $m^2/ \mathcal{M}^2 \approx 10^{-45}$, so that the relation between the NJL coupling $g$ and the UV cut-off $\mathcal{M}$ must be extremely fine-tuned in the range $1<g\mathcal{M}^2/2\pi^2<1 + 10^{-43}$, or $g \sim 2\pi^2\mathcal{M}^{-2}$, which can be safely neglected for current experimental limits. \\

\noindent
\textbf{E.} In four dimensions, if the kinetic term in the gravitational action (\ref{Lagrangian02}) is positive, as in the standard convention, the spacetime geometry is described by the Einstein-Hilbert action with positive cosmological constant. However, depending on the vacuum structure of the theory it might be worth considering the alternative where both $G$ and $\Lambda$ are negative, as in topologically massive gravity in three dimensions \cite{Carlip-Deser,Carlip:2008qh}. At any rate, the effective cosmological constant in the nontrivial vacuum should be given by $\Lambda_{Eff} =\Lambda + 2i\mu \langle \bar{\psi}\psi \rangle + g^{2}[\langle(\overline{\psi}\Gamma_5 \psi)^2\rangle - \langle(\overline{\psi}\psi)^2\rangle]$. It would be premature to claim something about the sign of $\Lambda_{Eff}$, especially in view of the fine tuning between $g$, $G$, $\Lambda$ and the cut-off $\mathcal{M}$.\\

\noindent
\textbf{F.} The gravitational Lagrangian is a particular combination of the three Lovelock terms that occur in 4D that has the form of the Pfaffian of the (A)dS curvature. This combination can also be viewed as the gravitational analogue of Born-Infeld electrodynamics \cite{BTZ}, and although the Gauss-Bonnet term has no affect on the field equations and hence is usually ignored, it can give a significant contribution to the global charges of the theory, and acts as a regulator that renders the charges well defined and finite in the presence of nontrivial asymptotics \cite{ACOTZ-1,ACOTZ-2}. It is therefore an interesting bonus of the model that the gravitational action is regularized by construction and no ad hoc counterterms are necessary to correctly define its thermodynamics.\\

\noindent
\textbf{G.} Even as an effective low energy model, a healthy theory should have a well defined (stable) ground state, a vacuum around which it would make sense to expand perturbatively to study the quantum features of the theory (Killing spinors, BPS vacua).  A vacuum without fermions (trivial vacuum, $\psi=0$) would be invariant under supersymmetry provided $\delta\psi=\nabla \lambda =0$, which means that $\lambda$ must be a covariantly constant (Killing) spinor. The number of linearly independent, globally defined solutions of this equation characterizes the residual supersymmetries of a particular background configuration. Such backgrounds have been studied in three and higher dimensions and a number of nontrivial BPS backgrounds are known \cite{Coussaert-Henneaux,Romans,Caldarelli-Klemm}. In the recent article \cite{TG-K}, the idea of replacing the Rarita-Schwinger field by a composite in an analogous manner to the one presented here, was also explored.\\

\noindent
\textbf{H.} In $2n+1$ dimensions, the CS form is the natural generalization of the construction in Section \textbf{3.1}. Clearly this is not the only option since a YM-term can always be included. However, the gauge symmetry of the action will be different if the Lagrangian is purely CS or YM. Restricting the analysis to CS forms only, the action can be expected to be invariant under the entire bosonic sector $G$ of the super-gauge group. All CS theories have a sector of solutions that is locally flat ($\mathbb{F}=0$) and with the fermions switched off. This is a maximally symmetric background with no propagating degrees of freedom \cite{Review}, which in $D=3$ is all there is. Those configurations enjoy full local supersymmetry as well, whereas a generic background in a different sector would not necessarily admit solutions of the projection constraints (\ref{epsilon-beta}), and would therefore not be SUSY-invariant in this sense.\\

\noindent
\textbf{I.} The situation for $D=2n$ is similar to the $D=4$ case, since the Chern-Weil theorem applies in general for $2n$-forms constructed out of $G$-connections: all $G$-invariant $2n$-forms are characteristic classes. The YM action can obviously be made to be invariant under $G_0 \times$[Lorentz], but that means that it is not off-shell SUSY-invariant. Moreover, in $2n$ dimensions it is unclear what kind of kinetic terms and couplings will be found for the gravitational and fermionic fields.

\section{Summary}       
The three- and four-dimensional models outlined above can be viewed as modeling the low energy limit of the standard model (QED), plus gravitation. The relation between these systems and supersymmetry is indirect and is reflected on the particular form of the field multiplets ($\psi$, $A_\mu$, $\omega^{ab}_\mu$, $e^a_\mu$) required by the superalgebra, and the specific couplings among these fields. The construction is characterized by the following features:\\
\noindent
$\bullet$ The representation is such that the fields are packaged into a connection one-form. Some features of standard supersymmetry are recovered --restricted multiplets of fields, reduced number of free parameter in the action, the need to include gravity in order to have the superalgebra acting locally. Other features of standard SUSY are not found: there is no matching of bosonic and fermionic degrees of freedom (no SUSY partners with equal quantum number except for the spin); no mass degeneracies: bosons remain massless, fermions acquire mass from couplings; bosons are gauge connections, fermions form conserved currents.

\noindent
$\bullet$ Including $s=1/2$ fermions in the superconnection requires the introduction of a metric structure ($e^a_\mu$), and the closure of the SUSY algebra requires the Lorentz group, which brings in the spin connection $\omega^{ab}$. Consequently the theory incorporates gravity in a natural manner: gravitation can be viewed a necessary consequence of having fermionic matter in nature. However, unlike standard local SUSY (supergravity) this theory has no spin-3/2 fields.

\noindent
$\bullet$ The restriction to $s=1/2$ requires projecting out the $s=3/2$ components generated by supersymmetry, a condition satisfied on locally flat ($\mathbb{F}=0$) backgrounds. Local flatness is satisfied by classical vacua in odd dimensions, but is expected to hold only approximately in even dimensions. For $D=2n+1$, the SUSY parameter $\epsilon$ is a spinor field whose form --if it exists-- depends on the background defined by the bosonic sector of the theory. Although the SUSY parameter $\epsilon$ is not constant, the symmetry does not correspond to a gauge invariance independent of the bosonic gauge field configurations. For $D=2n$, the AdS symmetry is broken at the level of the action by the fact that there are no $SO(2n,1)$- (or $SO(2n-1,2)$)-invariant tensors. Since the AdS symmetry is broken, supersymmetry is also necessarily broken\footnote{There can be accidents in some dimensions where other options exist for particular choices of fermionic representations such that $\{\mathbb{Q}, \overline{\mathbb{Q}}\}$ does not contain generators of AdS boosts \cite{TrZ2}.}.

\section*{Appendix A. Conventions}    
\textbf{Lorentz Group invariant tensors}\\
The signature we choose is such that the tangent space metric is $\eta_{ab}=diag(-1,1,1,1)$; the tangent space Levi-Civita invariant tensor of the Lorentz group is defined as
\[
\epsilon_{abcd}=\left\{\begin{array}{r l} 0 & \mbox{if any two indices repeat} \\
+1 & \mbox{even permutation of 0123} \\
-1 & \mbox{odd permutation of 0123}
\end{array}\right.
\]
so that, in particular $\epsilon_{0123}=+1=-\epsilon^{0123}$.\\

\noindent
\textbf{Levi-Civita tensor on a coordinate basis of the base manifold}\\
The alternating symbols in the base manifold ($\varepsilon$) are related to those on the tangent space ($\epsilon$) through
\begin{eqnarray}\nonumber
\varepsilon_{0123}=\epsilon_{0123}=+1 \, , \; \; \; \;  \varepsilon^{0123}=|e|^{-2}\epsilon^{0123}=-|e|^{-2}=|g|^{-1} \, , \\
e^a_\mu e^b_\nu e^c_\lambda e^d_\rho\epsilon_{abcd} =|e| \varepsilon_{\mu \nu \lambda \rho}\, ,\; \;  E_a^\mu E_b^\nu E_c^\lambda E_d^\rho\epsilon^{abcd} =|e| \varepsilon^{\mu \nu \lambda \rho} =|E|^{-1} \varepsilon^{\mu \nu \lambda \rho} \,   , \;
\nonumber
\end{eqnarray}
With these definitions, the volume form is
\[dx^\mu dx^\nu dx^\lambda dx^\rho=-|e|^2\varepsilon^{\mu \nu \lambda \rho} d^4x=-|e|E_a^\mu E_b^\nu E_c^\lambda E_d^\rho\epsilon^{abcd} d^4x \; ,
\]
and hence the oriented volume form is $e^0 e^1 e^2 e^3=|e|d^4x$. Also, if $\sigma=\frac{1}{2}\sigma_{\mu \nu}dx^\mu dx^\nu$ is a two-form, its Hodge-dual is
\[ ^*\sigma = \frac{1}{4}|e|\varepsilon_{\mu \nu \alpha \beta} \sigma^{\alpha \beta}dx^\mu dx^\nu\, .
\] \\

\noindent
\textbf{Dirac matrices}\\
The $\Gamma$-matrices are in a $4\times4$ spinorial-representation of the Clifford algebra $\{\Gamma^a,\Gamma^b\}=2 \eta^{ab}$, and $\Gamma^{ab} = \frac{1}{2} [\Gamma^a,\Gamma^b]$. The indices of the tangent space $a,b$ take the values 0,1,2 and 3. Consistently with the choice of signature, we take
\begin{eqnarray}
\Gamma^0=-\Gamma_0, \quad (\Gamma^0)^2 = (\Gamma_5)^2 =-1\,, \quad \mbox{where} \quad \Gamma_5 =\Gamma^0 \Gamma^1 \Gamma^2 \Gamma^3\, .
\end{eqnarray}
From this, a number of useful identities follow:
\begin{equation*}
\begin{array}{lrcl}
\textbf{I} & \Gamma_5\Gamma_a\Gamma_b\Gamma_c &=&\Gamma_5[\eta_{ab}\Gamma_c-\eta_{ac}\Gamma_b+\eta_{bc}\Gamma_a] + \epsilon_{abcd}\Gamma^d\\
\textbf{II} & \Gamma_5\Gamma_a\Gamma_b\Gamma_c \Gamma_d & =& \mbox{I} \epsilon_{abcd} + \Gamma_5 [\eta_{ab}\eta_{cd}- \eta_{ac} \eta_{bd}+ \eta_{ad}\eta_{bc}]\\
& &  &  +\Gamma_5[\eta_{ab}\Gamma_{cd} -\eta_{ac}\Gamma_{bd} +\eta_{ad}\Gamma_{bc} +\eta_{bc}\Gamma_{ad} -\eta_{bd}\Gamma_{ac} +\eta_{cd}\Gamma_{ab}]\\
\textbf{III} & \Gamma_a\Gamma_b\Gamma_c &=&  \eta_{ab} \Gamma_c -\eta_{ac} \Gamma_b +\eta_{bc} \Gamma_a -\epsilon_{abcd}\Gamma_5\Gamma^d \;.
\end{array}
\end{equation*}

\noindent
\textbf{Slashed notation}\\
Gamma matrices can be used to write Lorentz tensors in a spinorial basis, which is convenient sometimes when working with spin-1/2 fields. In our case, we have defined $\slashed{e} \equiv e^a \Gamma _a=\Gamma_\mu dx^\mu$ and $\slashed{\omega} \equiv\frac{1}{2}\omega^{ab} \Gamma _{ab}$. The covariant derivative of a spinor $\xi$ in the Lorentz connection becomes
\[
D\xi=d\xi + \frac{1}{2}\slashed{\omega} \xi, \quad \quad DD\xi= \slashed{R} \xi,
\]
where
\[
\slashed{R}=\frac{1}{2}R^{ab} \Gamma _{ab} = \left[d\slashed{\omega}+\frac{1}{2}\slashed{\omega}\slashed{\omega}\right]\; .
\]
If $\slashed{M}=\frac{1}{p!}dx^{\mu_{1}}\wedge\ldots\wedge dx^{\mu_{p}}M^{a_{1}\ldots a_{r}}_{\mu_{1}\ldots\mu_{p}}\Gamma_{a_{1}\ldots a_{r}}$ is a p-form spinorial tensor, then its covariant derivative reads
\[
D\slashed{M}=d \slashed{M} + \frac{1}{2}( \slashed{\omega} \slashed{M} -(-1)^p \slashed{M} \slashed{\omega}),
\]
which verifies Leibnitz's rule, $D(\slashed{M}\xi)=(D\slashed{M})\xi + (-1)^p \slashed{M} D\xi$.\\

\section*{Appendix B. Supersymmetric extension of $SO(3,1)\times U(1)$}   
The generators $\mathds{J}_a$ and $\mathds{J}_{ab}$ form the 4D (a)dS algebra
\begin{eqnarray} \label{J-J}
&[\mathds{J}_a,\mathds{J}_b]=s^2\mathds{J}_{ab}\,, \quad [\mathds{J}_a,\mathds{J}_{bc}]= \eta_{ab}\mathds{J}_c-\eta_{ac}\mathds{J}_b\,,&\\
&[\mathds{J}_{ab},\mathds{J}_{cd}]=\eta_{ad}\mathds{J}_{bc}-\eta_{ac}\mathds{J}_{bd}+ \eta_{bc}\mathds{J}_{ad} - \eta_{bd}\mathds{J}_{ac}\, ,
\end{eqnarray}
which corresponds to anti-de Sitter ($so(3,2)$) for $s=1$ and to de Sitter ($so(4,1)$) for $s=i$.  The supercharge $\mathds{Q}$ belongs to a spin 1/2 representation, that is
\begin{eqnarray} \label{J-Q}
& [\mathds{J}_a, \mathds{Q}^\alpha]=-\frac{s}{2}{ (\Gamma_a)^\alpha}_{\,\beta}\mathds{Q}^\beta, \quad    [\mathds{J}_a, \overline{\mathds{Q}}_\alpha]=\frac{s}{2}{\overline{\mathds{Q}}_\beta (\Gamma_a)^\beta}_{\,\alpha} ,\\
& [\mathds{J}_{ab}, \mathds{Q}^\alpha]=-\frac{1}{2}{ (\Gamma_{ab})^\alpha}_{\,\beta}\mathds{Q}^\beta, \quad [\mathds{J}_{ab}, \overline{\mathds{Q}}_\alpha]=\frac{1}{2}{\overline{\mathds{Q}}_\beta (\Gamma_{ab})^\beta}_{\,\alpha} .&
\label{J-Q-2}
\end{eqnarray}
Since $\mathds{Q}$ is complex, it has the following commutators with the $U(1)$ generator
\begin{equation} \label{K-Q}
[\mathds{K},\mathds{Q}^\alpha] = i\mathds{Q}^\alpha\, ,  \quad \qquad  [\mathds{K},  \overline{\mathds{Q}}_\alpha] = -i\overline{\mathds{Q}}_\alpha \;.
\end{equation}
The $\mathbf{\mathfrak{osp}(4|2)} \sim \mathbf{\mathfrak{usp}(2,2|1)}$ superalgebra is completed by the anticommutators of  supercharges,
\begin{equation}
\{\mathds{Q}^\alpha , \overline{\mathds{Q}}_\beta \}=-\frac{i}{s}(\Gamma^a)^{\alpha}_{\beta} \mathds{J}_a +\frac{i}{2} (\Gamma^{ab})^{\alpha}_{\beta}\mathds{J}_{ab} -\delta^{\alpha}_{\beta}\mathds{K} \, ,
\label{Q-Q}
\end{equation}
together with the trivial anticommutators $\{\overline{\mathds{Q}}_\alpha,\overline{\mathds{Q}}_\beta\}=0 = \{\mathds{Q}^\alpha,\mathds{Q}^\beta\}$. An explicit $6\times 6$ representation for the supercharges is the following
\begin{eqnarray}
(\mathds{Q}^\alpha)^A_{\,B} &=&- i \left[\begin{array}{c|cc}
0_{4\times4} & \begin{array}{c}
0\\
0\\
0\\
0
 \end{array} & C^{\alpha A} \\
\hline
\delta^{\alpha} _B& 0& 0 \\
0\; 0 \; 0\; 0\; & 0 & 0
\end{array}\right] \,  =-i(\delta^A_5 \delta^\alpha_B+ C^{\alpha A} \delta^6_B)\, \\
(\overline{\mathds{Q}}_\alpha)^A_{\,B} &=&\; \;  \left[\begin{array}{c|cc}
0_{4\times4} & \delta^A_\alpha & \begin{array}{c}
0\\
 0\\
 0\\
 0
 \end{array}\\
\hline
0\; 0 \; 0\; 0\; & 0 & 0 \\
C_{\alpha B} & 0& 0
\end{array}\; \; \; \right] \,   =\delta^A_\alpha \delta^5_B+\delta^A_6 C_{\alpha B}\, ,
\end{eqnarray}
where $C_{\alpha \beta}=-C_{\beta \alpha}$ is the conjugation matrix, and $C^{\alpha \beta}$ is its inverse.
In this representation, the $U(1)$ and (A)dS generators are
\begin{equation}
 \quad {(\mathds{K})^A}_B=\, \, \, \, \, \, \left[\begin{array}{c|cc}
 0_{4\times4} & \begin{array}{c}
 0\\
 0\\
 0\\
 0
 \end{array} & \begin{array}{c}
 0\\
 0\\
 0\\
 0
 \end{array}\\
\hline
0\; 0 \; 0\; 0\; & i & 0 \\
0\; 0 \; 0\; 0\; & 0 & \!\!\! -i
\end{array}\right] \, =i(\delta^A_5 \delta^5_B-\delta^A_6 \delta^6_B),
\end{equation}
\begin{eqnarray}
\quad {(\mathds{J}_a)^A}_B &=&\left[\begin{array}{c|cc}
 \frac{s}{2}\Gamma_a & \begin{array}{c}
 0\\
 0\\
 0\\
 0
 \end{array} & \begin{array}{c}
 0\\
 0\\
 0\\
 0
 \end{array}\\
\hline
0\; 0 \; 0\; 0\; & 0 & 0 \\
0\; 0 \; 0\; 0\; & 0 & 0
 \end{array}\right] \, =\frac{s}{2}{(\Gamma_a)^\alpha}_\beta \delta^A_\alpha \delta ^\beta_B\,   \\
\quad {(\mathds{J}_{ab})^A}_B&=&\left[\begin{array}{c|cc}
 \frac{1}{2}\Gamma_{ab} & \begin{array}{c}
 0\\
 0\\
 0\\
 0
 \end{array} & \begin{array}{c}
 0\\
 0\\
 0\\
 0
 \end{array}\\
\hline
0\; 0 \; 0\; 0\; & 0 & 0 \\
0\; 0 \; 0\; 0\; & 0 & 0
\end{array}\right] \,  =\frac{1}{2}{(\Gamma_{ab})^\alpha}_\beta \delta^A_\alpha \delta ^\beta_B\,
\end{eqnarray}
\\

\textbf{Acknowledgments.}
Enlightening discussions with A.~Balachandran, N.~Boulanger, F.~Canfora, G.~Catren, R.M.~Fernandes, G.~Giribet, M.~Hassa\"{\i}ne, J.~ Helayel-Neto, A.~Iorio, V.P.~Nair, H.~Nicolai, D.~Ponomarev, E.~Rodr\'{\i}guez, K.~Siampos, P.~Sundell, P.K.~Townsend, R.~Troncoso, and M.~Valenzuela are warmly acknowledged. We also thank A.~Chamseddine for bringing to our attention Refs. \cite{Chamseddine}. P.P. thanks the Physique Th\'{e}orique et Math\'{e}matique group of the Universit\'{e} Libre de Bruxelles, the International Solvay Institutes and the Faculty of Mathematics and Physics of Charles University in Prague for the kind hospitality during the development of this work,which was partially funded through a travel grant from Universidad Andr\'{e}s Bello's Vicerrector\'{i}a de Investigaci\'on y Doctorado. This work was supported by FONDECYT grant 1140155 and Becas Chile 74130061. The Centro de Estudios Cient\'{\i}ficos (CECS) is funded by the Chilean Government through the Centers of Excellence Base Financing Program of CONICYT-Chile.



\begin{thebibliography}{99}
\bibitem{Martin} S.~P.~Martin, \textit{A Supersymmetry primer}, in Perspectives on Supersymmetry II. Advanced Series on Directions in High Energy Physics, Vol. 18. Edited by Gordon L. Kane. World Scientific, Singapore, 1998 [hep-ph/9709356].

\bibitem{Murayama} H.~Murayama, \textit{Supersymmetry phenomenology}, in proceedings of ICTP-Particle Physics 1999 Summer School. G. Senjanovic and A.Yu. Smirnov, editors. World Scientific (Singapore, 2000).  [hep-ph/0002232].

\bibitem{Chatrchyan:2011zy} S.~Chatrchyan {\it et al.}  [CMS Collaboration], \textit{Search for Supersymmetry at the LHC in Events with Jets and Missing Transverse Energy}, Phys.\ Rev.\ Lett.\ {\bf 107} (2011) 221804 [arXiv:1109.2352 [hep-ex]].

\bibitem{Shifman} M.~Shifman, \textit{Frontiers Beyond the Standard Model: Reflections and Impressionistic Portrait of the conferernce}, Mod.\ Phys.\ Lett.\ A {\bf 27}, 1230043 (2012) [arXiv:1211.0004 [physics.pop-ph]].

\bibitem{Wess-Bagger} J.~Wess and J.~Bagger, \textit{Supersymmetry and Supergravity}, Princeton University Press (Princeton, 1992).

\bibitem{Freund} P.~G.~O.~Freund, \textit{Introduction to Supersymmetry}, Cambridge University Press (Cambridge, U.K., 1986).

\bibitem{Strings} M.~B.~Green, J.~H.~Schwarz and E.~Witten, Superstring Theory, Cambridge Monographs on Mathematical Physics, Cambridge University Press (Cam- bridge, UK, 1987).

\bibitem{Popper}K.~Popper, The Logic of Scientific Discovery, Hutchinson (London, 1959).

\bibitem{deWit} B.~de Wit and I.~Herger, \textit{Anti-de Sitter supersymmetry}, Lect.\ Notes Phys.\  {\bf 541}, 79 (2000) [hep-th/9908005].

\bibitem{Jackson-Okun} \textit{Historical roots of gauge invariance}, Rev.\ Mod.\ Phys.\  {\bf 73}, 663 (2001) [hep-ph/0012061].

\bibitem{Chamseddine} A.~Chamseddine, \textit{Massive Supergravity from Nonlinear Realization of Orthosymplectic Gauge Symmetry and Coupling to (Spin 1/2, Spin 1) Multiplet}, Nucl.\ Phys.\ B {\bf 131} (1977) 494 ; \textit{Topological gravity and supergravity in various dimensions}, Nucl.\ Phys.\ B {\bf 346} (1990) 213. A.~H.~Chamseddine and P.~C.~West, \textit{Supergravity as a Gauge Theory of Supersymmetry}, Nucl.\ Phys.\ B {\bf 129} (1977) 39.

\bibitem{BTrZ} M.~Ba\~nados, R.~Troncoso and J.~Zanelli, \textit{Higher dimensional Chern-Simons supergravity}, Phys.\ Rev.\ D {\bf 54}, 2605 (1996) [gr-qc/9601003].

\bibitem{TrZ1} R.~Troncoso and J.~Zanelli, \textit{New gauge supergravity in seven and eleven-dimensions}, Phys.\ Rev.\ D {\bf 58}, 101703 (1998) [hep-th/9710180].

\bibitem{TrZ2} R.~Troncoso and J.~Zanelli, \emph{Gauge Supergravities for all Odd Dimensions}, Int. Jour. Theor. Phys. \textbf{38}, 1181-1206 (1999) [hep-th/9807029].

\bibitem{HTrZ} M.~Hassaine, R.~Troncoso and J.~Zanelli, \textit{Poincar\'e invariant gravity with local supersymmetry as a gauge theory for the M-algebra}, Phys.\ Lett.\ B {\bf 596} (2004) 132 [hep-th/0306258].

\bibitem{AVZ} P.~D.~Alvarez, M.~Valenzuela and J.~Zanelli, \textit{Supersymmetry of a different kind}, JHEP {\bf 1204} (2012) 058 [arXiv:1109.3944 [hep-th]].

\bibitem{WZ} J.~Wess and B.~Zumino, \textit{Supergauge Transformations in Four-Dimensions}, Nucl.\ Phys.\ B {\bf 70}, 39 (1974).

\bibitem{Nakahara} M.~Nakahara, \textit{Geometry, topology and physics}, Taylor and Francis (Boca Raton, USA, 2003).

\bibitem{CJS} E.~Cremmer, B.~Julia and J.~Scherk, \textit{Supergravity Theory in Eleven-Dimensions}, Phys.\ Lett.\ B {\bf 76}, 409 (1978).

\bibitem{Schonfeld}  J.~F.~Schonfeld, \textit{A Mass Term for Three-Dimensional Gauge Fields}, Nucl.\ Phys.\ B {\bf 185} (1981) 157.

\bibitem{DJT} S.~Deser, R.~Jackiw and S.~Templeton, \textit{Three-Dimensional Massive Gauge Theories}, Phys.\ Rev.\ Lett.\  {\bf 48} (1982) 975; \textit{Topologically Massive Gauge Theories}, Annals Phys.\  {\bf 140} (1982) 372; [Erratum-ibid.\  {\bf 185} (1988) 406;  {\bf 281} (2000) 409]

\bibitem{Review} J.~Zanelli, \textit{Lecture notes on Chern-Simons (super-)gravities}. Second edition (February 2008), [hep-th/0502193].

\bibitem{MD-M} S.~W.~Mac Dowell and F.~Mansouri, \textit{Unified Geometric Theory of Gravity and Supergravity}, Phys.\ Rev.\ Lett.\  {\bf 38} (1977) 739; [Erratum-ibid.\  {\bf 38} (1977) 1376].

\bibitem{Townsend}P.~K.~Townsend, \textit{Small Scale Structure of Space-Time as the Origin of the Gravitational Constant}, Phys.\ Rev.\ D {\bf 15} (1977) 2795.

\bibitem{APRSZ}P.~D.~Alvarez, P.~Pais, E.~Rodriguez, P.~Salgado and J.~Zanelli, (in preparation).

\bibitem{Weyl} H.~Weyl, \textit{A Remark on the coupling of gravitation and electron}, Phys.\ Rev.\  {\bf 77} (1950) 699.

\bibitem{Miskovic-Z} O.~Miskovic and J.~Zanelli, \textit{On the negative spectrum of the 2+1 black hole}, Phys.\ Rev.\ D {\bf 79} (2009) 105011 [arXiv:0904.0475 [hep-th]].

\bibitem{EGMZ} J.~D.~Edelstein, A.~Garbarz, O.~Miskovic and J.~Zanelli, \textit{Stable p-branes in Chern-Simons AdS supergravities}, Phys.\ Rev.\ D {\bf 82} (2010) 044053 [arXiv:1006.3753 [hep-th]]; \textit{Geometry and stability of spinning branes in AdS gravity}, Phys.\ Rev.\ D {\bf 84} (2011) 104046 [arXiv:1108.3523 [hep-th]].

\bibitem{VKG} M.A.H.~Vozmediano, M.I.~Katsnelson and F.~Guinea, \textit{Gauge fields in graphene}, Phys.\ Rept.\  {\bf 496} (2010) 109.

\bibitem{MSZ} A.~Mesaros, D.~Sadri and J.~Zaanen1, \textit{Parallel Transport of Electrons in Graphene Parallels Gravity}, Phys.\ Rev.\ B {\bf 82} (2010) 073405 [arXiv:0909.2703 [cond-mat.mes-hall]].

\bibitem{Cvetic-Gibbons} M.~Cveti\'c and G.~WGibbons, \textit{Graphene and the Zermelo Optical Metric of the BTZ Black Hole}, Annals Phys.\  {\bf 327} (2012) 2617 [arXiv:1202.2938 [hep-th]].

\bibitem{Iorio} A.~Iorio and G.~Lambiase, \textit{Quantum field theory in curved graphene spacetimes, Lobachevsky geometry, Weyl symmetry, Hawking effect, and all that}; [arXiv:1308.0265 [hep-th]].

\bibitem{Carroll}  S.~M.~Carroll and G.~B.~Field, \textit{Consequences of propagating torsion in connection dynamic theories of gravity}, Phys.\ Rev.\ D {\bf 50} (1994) 3867 [gr-qc/9403058].

\bibitem{Wise} D~K.~Wise, \textit{Symmetric space Cartan connections and gravity in three and four dimensions}, SIGMA {\bf 5} (2009) 080 [arXiv:0904.1738 [math.DG]]; \textit{MacDowell-Mansouri gravity and Cartan geometry}, Class.\ Quant.\ Grav.\  {\bf 27} (2010) 155010 [gr-qc/0611154].

 \bibitem{Catren} G.~Catren, \textit{Geometric foundations of classical Yang-Mills theory}, Stud.\ Hist.\ Philos.\ Mod.\ Phys.\  {\bf 39} (2008) 511.

\bibitem{AWZ} A. Anabal\'on, S. Willison, J. Zanelli, \textit{General relativity from a gauged WZW term}, Phys.\ Rev.\ D {\bf 75} (2007) 024009 [hep-th/0610136].

\bibitem{MPW} P. Mora, P. Pais, S. Willison, Gauged WZW models for space-time groups and gravitational actions, Phys.\ Rev.\ D {\bf 84} (2011) 044058 [arXiv:1107.0758 [hep-th]].

\bibitem{Nambu} Y.~Nambu and G.~Jona-Lasinio, Phys.\ Rev.\  {\bf 122} (1961) 345; Phys.\ Rev.\  {\bf 124} (1961) 246. See also, \textit{A `superconductor' model of elementary particles and its consequences}, in ``Broken Symmetry", Collected papers of Y. Nambu, edited by  T.~Eguchi and K.~Nishijima, World Scientific Series in 20th Century Physics - Vol. 13, Singapore, 1995.

\bibitem{Loewe:2013zaa} M.~Loewe, F.~Marquez and C.~Villavicencio, The nNJL model with a fractional Lorentzian regulator in the real time formalism, Phys.\ Rev.\ D {\bf 88} (2013) 056004 [arXiv:1307.6764 [hep-ph]].

\bibitem{Carlip-Deser} S.~Carlip, S.~Deser, A.~Waldron and D.~K.~Wise, T\textit{opologically Massive AdS Gravity}, Phys.\ Lett.\ B {\bf 666} (2008) 272 [arXiv:0807.0486 [hep-th]].

\bibitem{Carlip:2008qh} S.~Carlip, The Constraint Algebra of Topologically Massive AdS Gravity, JHEP {\bf 0810} (2008) 078 [arXiv:0807.4152 [hep-th]].

\bibitem{BTZ} M.~ Ba\~nados, C.~Teitelboim and J.~Zanelli, \textit{Lovelock-Born-Infeld theory of gravity}, in J.J. Giambiagi Festschrift, H. Falomir, et al (editors), World Scientific, Singapore, 1990.

\bibitem{ACOTZ-1} R.~Aros, M.~Contreras, R.~Olea, R.~Troncoso and J.~Zanelli, \textit{Conserved charges for gravity with locally AdS asymptotics}, Phys.\ Rev.\ Lett.\  {\bf 84} (2000) 1647 [gr-qc/9909015].

\bibitem{ACOTZ-2} R.~Aros, M.~Contreras, R.~Olea, R.~Troncoso and J.~Zanelli, \textit{Conserved charges for even dimensional asymptotically AdS gravity theories}, Phys.\ Rev.\ D {\bf 62} (2000) 044002 [hep-th/9912045].

\bibitem{Coussaert-Henneaux} O.~Coussaert and M.~Henneaux, \textit{Supersymmetry of the (2+1) black holes}, Phys.\ Rev.\ Lett.\  {\bf 72} (1994) 183 [hep-th/9310194].

\bibitem{Romans} L.~J.~Romans, \textit{Supersymmetric, cold and lukewarm black holes in cosmological Einstein-Maxwell theory}, Nucl.\ Phys.\ B {\bf 383} (1992) 395  [hep-th/9203018].

\bibitem{Caldarelli-Klemm} M.~M.~Caldarelli and D.~Klemm, \textit{Supersymmetry of Anti-de Sitter black holes}, Nucl.\ Phys.\ B {\bf 545} (1999) 434 [hep-th/9808097].

\bibitem{TG-K} A.~Torres-Gomez and K.~Krasnov, \textit{Fermions via spinor-valued one-forms}, Int.\ J.\ Mod.\ Phys.\ A {\bf 28} (2013) 24, 1350113. [arXiv:1212.3452].

\end{thebibliography}
\end{document}